\documentclass[useAMS,usenatbib,usegraphicx]{mn2e}
\usepackage{epsf}

\newcommand{\bm}[1]{\mbox{\boldmath$#1$}}

\title[Scatter and bias in weak lensing selected clusters]
{Scatter and bias in weak lensing selected clusters}

\author[T. Hamana et al.]
{Takashi Hamana$^{1}$, Masamune Oguri$^{2}$, Masato Shirasaki$^{3}$ and
  Masanori Sato$^{4}$\\ 
$^{1}$National Astronomical Observatory of Japan, Mitaka, Tokyo
  181-8588, Japan.\\
$^{2}$Kavli Institute for the Physics and Mathematics of the Universe,
  University of Tokyo, Chiba 277-8582, Japan\\
$^{3}$Department of Physics, Graduate School of Science, The
  University of Tokyo, Tokyo 113-0033, Japan\\
$^{4}$Department of Physics, Nagoya University, Nagoya 464-8602, Japan} 

\begin{document}

\date{\today}

\voffset- .5in

\pagerange{\pageref{firstpage}--\pageref{lastpage}} \pubyear{}

\maketitle

\label{firstpage}

\begin{abstract}
We examine scatter and bias in weak lensing selected clusters,
employing both an analytic model of dark matter haloes and numerical
mock data of weak lensing cluster surveys. We pay special attention 
to effects of the diversity of dark matter distributions within
clusters. We find that peak heights of the lensing convergence map
correlates rather poorly with the virial mass of haloes. The
correlation is tighter for the spherical overdensity mass with a higher
mean interior density (e.g., $M_{1000}$). We examine the dependence of
the halo shape on the peak heights, and find that the root-mean-square
scatter caused by the halo diversity scales linearly with the peak
heights with the proportionality factor of $0.1-0.2$. The noise
originated from the halo shape is found to be comparable to the source
galaxy shape noise and the cosmic shear noise. We find the significant
halo orientation bias, i.e., weak lensing selected clusters on average
have their major axes aligned with the line-of-sight direction, and that
the orientation bias is stronger for higher signal-to-noise ratio
($S/N$) peaks. We compute the orientation bias using an analytic
triaxial halo model and obtain results quite consistent with the
ray-tracing results. We develop a prescription to analytically compute
the number count of weak lensing peaks taking into account all the
main sources of scatters in peak heights. We find that the improved
analytic predictions agree well with the simulation results for high
$S/N$ peaks of $\nu\ga5$. We also compare the expected number count
with our weak lensing analysis results for 4~deg$^2$ of
Subaru/Suprime-Cam observations and find a good agreement. 
\end{abstract}

\begin{keywords}
cosmology: theory 
--- dark matter 
--- galaxies: clusters: general 
--- gravitational lensing
\end{keywords}

%%%%%%%%%%%%%%%%%%%%%%%%%%%%%%%%%%%%%%%%%%%%%%%%%%%%%%%%%%%
%%%%%%%%%%%%%%%%%%%%%%%%%%%%%%%%%%%%%%%%%%%%%%%%%%%%%%%%%%%
\section{Introduction}
\label{sec:intro}
%%%%%%%%%%%%%%%%%%%%%%%%%%%%%%%%%%%%%%%%%%%%%%%%%%%%%%%%%%%
%%%%%%%%%%%%%%%%%%%%%%%%%%%%%%%%%%%%%%%%%%%%%%%%%%%%%%%%%%%

Clusters of galaxies have been playing an important role in the field
of cosmology. For instance, number counts of clusters have placed
useful constraints on cosmological parameters
\citep[e.g.,][]{vikhlinin09b,mantz10,rozo10},  
and measurement of the dark matter distribution in clusters
is useful for testing dark matter models
\citep[e.g.,][]{dahle06,broadhurst08,mahdavi08,okabe08,oguri09,oguri10,oguri12,umetsu09,umetsu11,medezinski10,okabe10}. 
In most of these applications, the construction of homogeneous
cluster samples and understanding selection biases are of fundamental
importance, because an unknown bias can significantly affect the
interpretation of the data. 
 
Clusters of galaxies are identified by various techniques, including
detections of galaxy concentrations in optical data, extended
X-ray emissions \citep[e.g.,][]{Bohringeretal04,vikhlinin09a}, the
Sunyaev-Zel'dovich effect in the cosmic microwave background
\citep[e.g.,][]{Marriageetal11,Reichardtetal12}, and dark matter
concentrations in the weak lensing mass map \citep[e.g.,
][]{miyazaki02,miyazaki07,Wittmanetal06,gavazzi07,schirmer07,kubo09,bellagamba11,shan12,kurtz12}. 
Each methodology has its own advantage and disadvantage. Among others,
weak lensing technique is unique in the sense that it identifies
clusters by searching for high peaks in the weak lensing mass 
map, and thus does not rely on physical state of the baryonic
component \citep*{schneider96,hamana04,hennawi05,maturi05}.
Therefore, combining weak lensing cluster sample with samples selected
by another method is beneficial not only for various cosmological
applications but also for better understanding of cluster physics.

Recently \citet{shan12} reported 301 weak lensing high peaks located from
64~deg$^2$ data. Among those peaks, they confirmed 85 groups/clusters,
which is the largest weak lensing cluster samples constructed to date 
(see also \citet{miyazaki07} and \citet{schirmer07}).
Currently the size of weak
lensing selected cluster sample is mainly limited by areas of optical
imaging surveys. However, the situation is changing drastically in coming
decade because there are several ongoing or planned wide field deep
optical surveys including Dark Energy   
Survey\footnote{\tt{http://www.darkenergysurvey.org/}}, 
the KIlo-Degree Survey\footnote{http://www.astro-wise.org/projects/KIDS/}, 
Hyper Suprime-Cam\footnote{\tt{http://subarutelescope.org/Projects/HSC/}},  
and Large Synoptic Survey Telescope\footnote{\tt{http://www.lsst.org/lsst/}}.
Thus in the near future, weak lensing selected cluster catalogues
containing $\sim10,000$ clusters will be available. It is therefore
important to examine the selection function of the weak lensing
selected cluster samples in order to take full advantage of these
unique cluster catalogues.

The standard method for identifying clusters with weak lensing is to
search for high peaks in the weak lensing mass map generated from weak
lensing shear data with a carefully designed smoothing filter. 
Peaks above a given threshold are selected as cluster candidates.
The peak height, which plays a central role in this study, is primarily
determined by the mass and redshift of the dark matter halo of
clusters \citep[see e.g.,][]{hamana04}, but is also affected by
the following three effects. The first is the noise arising from
intrinsic shapes of source galaxies used for weak lensing shear
measurements (galaxy shape noise). The second is the projection of
structures along the line-of-sight. The third is the diversity of dark
matter distributions in individual haloes. These effects are important
in the sense that they induce the scatter and bias in the peak heights
and hence in the resulting weak lensing selected cluster sample. While
these effects were examined in literature to some extent
\citep*{hamana04,hennawi05,maturi05,TangFan2005,pace07,Fanetal2010,marian10,SchmidtRozo2011,dietrich12}, 
we revisit this problem with a particular emphasis on the scatter and
bias produced by the diversity of dark matter distributions, using both
analytic and numerical approaches.

The matter content of clusters is dominated by dark matter, which
accounts for $\sim 80$ per cent of the mass of the universe. The dark matter
distribution within dark matter haloes are investigated in detail using
$N$-body simulations, which found that the radial profile of the dark
matter halo is accurately described by an analytic form, the so-called
NFW profile \citep{navarro97}. In this model, the matter distribution
of haloes is characterized by two parameters, the mass and the
concentration parameter. While it is known that there is a mean
relationship between the mass and concentration parameter, the
relation involves a large scatter among different haloes
\citep[e.g.,][]{Bullocketal2001}. As pointed out by
\citet{KingMead2011}, the weak lensing peak height is sensitive to the
halo concentration. Another important prediction of the current
standard structure formation model is that the dark matter
distribution in clusters is not spherically symmetric but is highly
elongated \citep{JingSuto2002}. This non-sphericity of the dark matter
distribution is known to have large impact on weak lensing
measurements of clusters 
\citep{clowe04,oguri05,gavazzi05,TangFan2005,corless09,FerozHobson2012}.
Thus even the same mass haloes at the same redshift can produce
largely different peak heights depending on the halo concentration and
non-sphericity. This is exactly what we explore in this paper using a
large set of mock data generated from gravitational lensing
ray-tracing in $N$-body simulations. 

The structure of this paper is as follows. In Section~\ref{sec:theory}
we summarize the analytic description of cluster density distribution
adopted in this paper and also summarize basics of cluster finding with
weak lensing. We use a mock numerical simulation of weak lensing
cluster survey which is detailed in Section~\ref{sec:simulation}. 
In Section~\ref{sec:results} we investigate the scatter and bias in the
weak lensing selected clusters, paying special attention to the
influence of halo shape. We briefly compare the peak number count
derived from the numerical mock data with our weak lensing analysis
results of Subaru/Suprime-Cam observations in
Section~\ref{sec:scam}. Finally, summary and discussion are given in
Section~\ref{sec:summary}.   

Throughout this paper we adopt the cosmological model with  
the matter density $\Omega_M=0.238$, baryon density $\Omega_b=0.042$, 
cosmological constant $\Omega_\Lambda=0.762$, spectral index
$n_s=0.958$, the normalization of the matter fluctuation
$\sigma_8=0.76$, and the Hubble parameter $h=0.732$, which are the
best-fit cosmological parameters in the Wilkinson Microwave Anisotropy
Probe (WMAP) third-year results \citep{Spergel07}.

%%%%%%%%%%%%%%%%%%%%%%%%%%%%%%%%%%%%%%%%%%%%%%%%%%%%%%%%%%%
%%%%%%%%%%%%%%%%%%%%%%%%%%%%%%%%%%%%%%%%%%%%%%%%%%%%%%%%%%%
\section{Analytic models of dark matter haloes}
\label{sec:theory}
%%%%%%%%%%%%%%%%%%%%%%%%%%%%%%%%%%%%%%%%%%%%%%%%%%%%%%%%%%%
%%%%%%%%%%%%%%%%%%%%%%%%%%%%%%%%%%%%%%%%%%%%%%%%%%%%%%%%%%%

Let us first summarize the analytic models of dark matter distribution of
clusters of galaxies and its weak lensing properties.

%%%%%%%%%%%%%%%%%%%%%%%%%%%%%%%%%%%%%%%%%%%%%%%%%%%%%%%%%%
\subsection{Dark matter distribution within haloes}
\label{sec:halo-theory}
%%%%%%%%%%%%%%%%%%%%%%%%%%%%%%%%%%%%%%%%%%%%%%%%%%%%%%%%%%%

We adopt the NFW model for the dark halo density
profile which is given by \citep{navarro97},
%%%%%%%%%%%
\begin{equation}
\rho_{\rm NFW}(r)=\frac{\rho_s}{(r/r_s)(1+r/r_s)^2}.
\label{eq:nfw}
\end{equation}
%%%%%%%%%%%
This model characterizes the dark matter distribution by two
parameters: the density parameter $\rho_s$ and the scale radius $r_s$.
It is customary to re-characterizes it by related parameters, the
halo mass and the concentration parameter, which we define below.

The mass of haloes is not a uniquely defined quantity. One needs to
choose proper definitions of masses depending on the purposes
\citep{white01}. In theoretical and observations studies, the virial 
mass $M_{\rm vir}$, which is defined such that the average density
within the virial radius becomes equal to the nonlinear overdensity
$\Delta_{\rm vir}$ computed using the spherical collapse model \citep[see,
  e.g.,][]{nakamura97} times the mean matter density of the
universe, has often been adopted. For the NFW model, the virial mass
relates to $\rho_s$ and $r_s$ by 
%%%%%%%%%%%
\begin{equation}
\label{eq:mvir}
M_{\rm vir}=
\frac{4 \pi}{3} \Delta_{\rm vir}(z)\bar{\rho}_m(z)r_{\rm vir}^3
=
4 \pi \rho_s r_s^3 m_{\rm nfw}(c_{\rm vir}),
\end{equation}
%%%%%%%%%%%
where $c_{\rm vir}$ is the so-called concentration parameter defined
by
%%%%%%%%%%%
\begin{equation}
\label{eq:c-m}
c_{\rm vir}
\equiv \frac{r_{\rm vir}}{r_s}
=\frac{1}{r_s}\left[\frac{3M_{\rm vir}}
{4\pi\Delta_{\rm vir}(z)\bar{\rho}_m(z)}\right]^{1/3},
\end{equation}
%%%%%%%%%%%
and $m_{\rm nfw}(c_{\rm vir})$ defined by
%%%%%%%%%%%
\begin{equation}
m_{\rm nfw}(c_{\rm vir})\equiv\int_0^{c_{\rm vir}}
\frac{x}{(1+x)^2}dx=\ln(1+c_{\rm vir})-\frac{c_{\rm vir}}{1+c_{\rm vir}}.
\end{equation}
%%%%%%%%%%%
The concentration parameter is known to be correlated with the halo
mass and redshift. When necessary, we adopt the following relationship:
%%%%%%%%%%%
\begin{equation}
c_{\rm vir}(M_{\rm vir},z)=7.26\left(\frac{M_{\rm
    vir}}{10^{12}h^{-1} M_\odot}\right)^{-0.086}\left(1+z\right)^{-0.71},
\label{eq:concentation}
\end{equation}
%%%%%%%%%%%
which was derived from $N$-body simulations assuming the WMAP third year
results \citep{maccio08}, with the additional
redshift dependence based on the simulation result of \citet{duffy08}.

While the virial mass is a physically motivated definition of the halo
mass, one can define halo masses using arbitrary values of
overdensities. Specifically, one can use the spherical overdensity
mass $M_\Delta$ defined by the mass contained within a radius 
$r_\Delta$ inside of which the mean interior density is $\Delta$ times
the {\it critical} density\footnote{Some authors adopt a different
  definition in which the $\Delta$ is specified relative to the average
  background density $\bar{\rho}_m(z)$. Our $\Delta$ is $\Omega_m(z)$
  times theirs.}
%%%%%%%%%%%
\begin{equation}
\label{eq:mso}
M_\Delta=
\frac{4 \pi}{3} \Delta \rho_{cr}(z)r_\Delta^3
=
4 \pi \rho_s r_s^3 m_{\rm nfw}(c_\Delta),
\end{equation}
%%%%%%%%%%%
where
%%%%%%%%%%%
\begin{equation}
c_\Delta
\equiv \frac{r_\Delta}{r_s}
=\frac{1}{r_s}\left[\frac{3M_\Delta}
{4\pi\Delta \rho_{cr}(z)}\right]^{1/3}.
\end{equation}
%%%%%%%%%%%

In this paper, we also consider a triaxial halo model of
\citet{JingSuto2002} to investigate the effect of the halo
triaxiality.  In this model, the density profile given by
eq.~(\ref{eq:nfw}) is modified as
%%%%%%%%%%%
\begin{equation}
\rho_{\rm tri}(r)=\frac{\rho_{cd}}{(R/R_0)(1+R/R_0)^2},
\label{eq:tri}
\end{equation}
%%%%%%%%%%%
\begin{equation}
R^2\equiv c^2\left(\frac{x^2}{a^2}+\frac{y^2}{b^2}
+\frac{z^2}{c^2}\right)\;\;\;\;\;(a\leq b\leq c).
\label{eq:rdef}
\end{equation}
%%%%%%%%%%%
\citet{JingSuto2002} derived the probability distribution of axis
ratios $a/c$ and $b/c$ for haloes with a given mass and redshift from
large $N$-body simulations. The lensing properties of the triaxial
halo model were derived in \citet{oguri03} and \citet{oguri04},
including projections of triaxial haloes along arbitrary directions,
which we adopt in this paper. We study the effect of the halo
triaxility on the properties of weak lensing selected clusters using
the semi-analytic approach developed by \citet{ob09}. In this method,
we generate a catalogue of haloes according to the mass function as
well as axis ratio distributions of \citet{JingSuto2002}, and project
each halo along random direction to compute the lensing
properties. This allows us to generate a mock catalogue of weak
lensing selected clusters based on the triaxial halo model. 

%%%%%%%%%%%%%%%%%%%%%%%%%%%%%%%%%%%%%%%%%%%%%%%%%%%%%%%%%%
\subsection{Basics of weak lensing cluster finding}
\label{sec:lensing-theory}
%%%%%%%%%%%%%%%%%%%%%%%%%%%%%%%%%%%%%%%%%%%%%%%%%%%%%%%%%%%
Here we summarize equations for weak lensing cluster finding which are
directly relevant to the following analyses. For more detailed
descriptions, see 
\citet{schneider96,bartelmannetal2001,hamana04,hennawi05,maturi05}.

Let us first defined the weak lensing mass map which is the smoothed
lensing convergence field ($\kappa$):
%%%%%%%%%%%
\begin{equation}
{\cal K }(\bm{\theta}) 
=\int d^2\bm{\phi}~ \kappa(\bm{\phi}-\bm{\theta}) U(|\bm{\phi}|),
\end{equation}
%%%%%%%%%%%
where $U$ is the filter function to be specified below.
The same quantity is obtained from the shear data by
%%%%%%%%%%%
\begin{equation}
\label{eq:shear2kap}
{\cal K }(\bm{\theta}) 
=\int d^2\bm{\phi}~ \gamma_t(\bm{\phi}:\bm{\theta}) Q(|\bm{\phi}|),
\end{equation}
%%%%%%%%%%%
where $\gamma_t(\bm{\phi}:\bm{\theta})$ is the tangential component of
the shear at position $\bm{\phi}$ relative to the point $\bm{\theta}$,
and $Q$ relates to $U$ by 
%%%%%%%%%%%
\begin{equation}
Q(\theta) 
=\int_0^\theta d\theta'~\theta' U(\theta')-U(\theta).
\end{equation}
%%%%%%%%%%%
We consider $Q$ with a finite extent; in this case, one finds
%%%%%%%%%%%
\begin{equation}
U(\theta) 
=2 \int_\theta^{\theta_o} d\theta'~ {{Q(\theta')} \over {\theta'}}
-Q(\theta),
\end{equation}
%%%%%%%%%%%
where $\theta_o$ is the outer boundary of the filter.
Note that this is equivalent to set $U$ a finite compensated filter; 
that is, $\int^{\theta_o} d\theta~ \theta ~U(\theta) = 0$ and $U(\theta) =
0$ for $\theta > \theta_o$.

The basic idea of weak lensing cluster finding is to first construct
a weak lensing mass  map by applying eq.~(\ref{eq:shear2kap}) to shear
data, then to search for high peaks in the map which are plausible
candidates of massive clusters. The root-mean-square (rms) noise
coming from intrinsic ellipticity of galaxies (which we call the
galaxy shape noise) is evaluated by \citep{schneider96},
%%%%%%%%%%%
\begin{equation}
\sigma_{\rm shape}^2 
={{\sigma_e^2}\over{2n_g}} \int_0^{\theta_o} d\theta~\theta~Q^2(\theta),
\end{equation}
%%%%%%%%%%%
where $\sigma_e$ is the rms value of intrinsic ellipticityies of galaxies 
and $n_g$ is the number density of galaxies.
Throughout this paper, we take $\sigma_e=0.4$ and $n_g=30~{\rm
  arcmin}^{-2}$, which resembles the shape noise expected for the Hyper
Suprime-Cam survey. The signal-to-noise ratio ($S/N$) of the weak
lensing map is defined by the ratio between the peak height and
$\sigma_{\rm shape}$,   
%%%%%%%%%%%
\begin{equation}
\nu = {{\cal K } \over {\sigma_{\rm shape}}}.
\end{equation}
%%%%%%%%%%%

%
%%%%% Fig-1
%
\begin{figure}
\begin{center}
 \includegraphics[width=84mm]{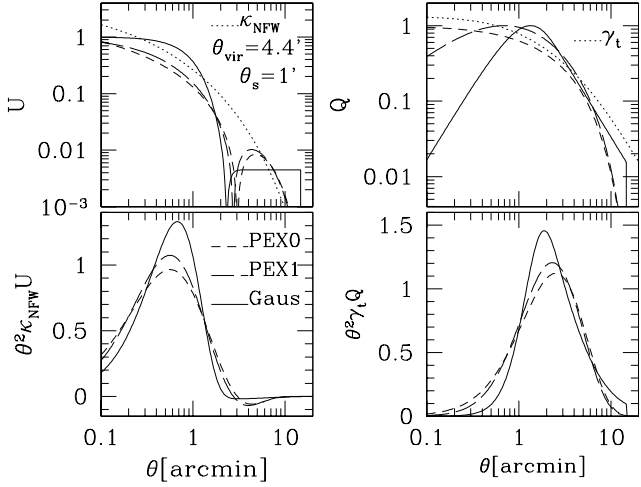}
\end{center}
\caption{Three window functions considered in this paper are plotted.
{\it Top left}: Absolute value of $U$ is shown, with negative $U$
being plotted by thin lines. The normalization is arbitrary. 
The convergence profile of the smoothly truncated NFW halo 
($M_{\rm vir}=10^{14}h^{-1} M_\odot$, $z_l=0.3$, $z_s=1$)
\citep{OguriHamana2011} is also plotted as the dotted line. 
{\it Bottom left}: $\theta^2 \kappa_{\rm NFW} U $, which is the
contribution to the ${\cal K}$ from the logarithmic $\theta$ interval,
is plotted.  
{\it Top right}: The filter $Q$. The normalization is arbitrary. 
The tangential shear profile of the smoothly truncated NFW halo
is also plotted.
{\it Bottom right}: $\theta^2 \gamma_t Q $, which is the contribution
to the ${\cal K}$ from the logarithmic $\theta$ interval, is plotted. 
\label{fig1.ps}}
\end{figure}

In this paper we consider the following three filter functions.
One is the truncated Gaussian (for $U$),
%%%%%%%%%%%
\begin{equation}
Q_{G}(\theta) 
={1 \over {\pi \theta^2}}
\left[1-\left(1+{{\theta^2}\over {\theta_G^2}}\right)
\exp\left(-{{\theta^2}\over {\theta_G^2}}\right)\right],
\end{equation}
%%%%%%%%%%%
for $\theta < \theta_o$ and $Q_G=0$ elsewhere. We take $\theta_G=1'$,
which is the value also adopted in \citet{hamana04}, and $\theta_o=15'$. 
The others filters have the following functional form, consisting of
the power-law with outer exponential-cutoff,  
%%%%%%%%%%%
\begin{equation}
Q_{PEXn}(\theta) 
={(\theta/\theta_f)^n \over {\theta_f^2(1+a \theta /\theta_f)^{(2+n)}}}
\exp\left(-{{\theta^2}\over {2 \theta_e^2}}\right).
\end{equation}
%%%%%%%%%%%
We consider two cases: $(n,a)=(0,0.25)$ and $(1,0.7)$ which we call
PEX0 and PEX1, respectively.
The former mimics the filter function proposed by \citet{maturi05}
designed for maximizing the $S/N$ of weak lensing peak by the NFW halo
relative to noises coming from galaxy shape and cosmic structures,
whereas the latter mimics the one proposed by \citet{hennawi05} which
has a similar shape at the outer region but has less power on the inner region
for suppressing the galaxy shape noise. We take $\theta_f=1'$,
$\theta_e=5'$ and $\theta_o=15'$ for both the cases, which are chosen 
so as to maximize the $S/N$ for a cluster at $z=0.3$ with $M_{\rm
  vir}=10^{14}h^{-1} M_\odot$. Note that the filter scales do not need to
be fixed but in general can be varied to act as a {\it matched filter}
\citep[e.g.,][]{hennawi05,marian10}. In this paper we do not vary the
filter scales because it is beyond the scope of this paper. The
shapes of the filter functions are plotted in two top panels of
Fig.~\ref{fig1.ps}.  

%
%%%%% Fig-2
%
\begin{figure}
\begin{center}
 \includegraphics[width=80mm]{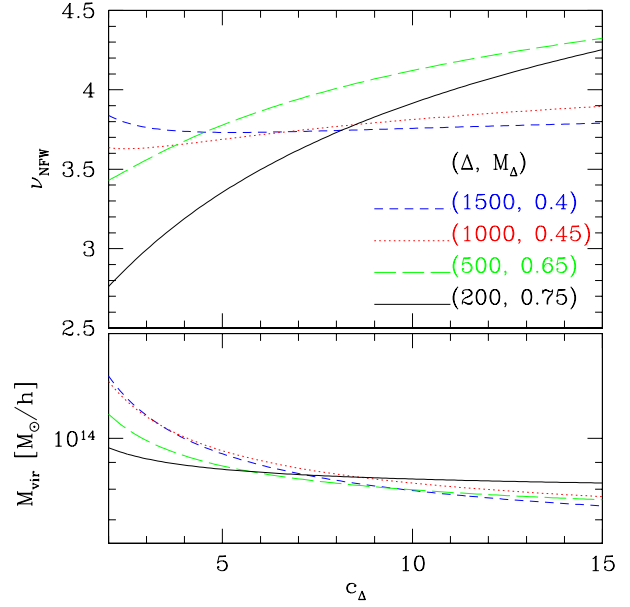}
\end{center}
\caption{{\it Top}: The peak $S/N$ expected for NFW halo as a function of
  the concentration parameter. Different lines show different
  spherical overdensity mass shown in unit of
  $10^{14}h^{-1} M_{\odot}$. The lens and source redshift are set to be 0.3
  and 1, respectively. 
{\it Bottom}: Corresponding virial mass is plotted.
\label{fig2.ps}}
\end{figure}

The weak lensing peak height for the NFW halo is computed by
%%%%%%%%%%%
\begin{equation}
\label{eq:kappa_nfw}
{\cal K }_{\rm NFW}
=2\pi \int d\phi~\phi~ \kappa_{\rm NFW}(\phi) U(\phi),
\end{equation}
%%%%%%%%%%%
where $\kappa_{\rm NFW}$ is the convergence profile from NFW halo for
which we take the smoothly truncated NFW profile 
\citep[see ][for analytic expressions]{OguriHamana2011,baltz09}.
It is seen from bottom panels
of Fig.~\ref{fig1.ps} that the most of the
contribution to ${\cal K}_{\rm NFW}$ comes from the matter within the scale
radius or from shear data on scales $\theta_s<\theta<\theta_{\rm vir}$.
We denote the $S/N$ expected from NFW halo by 
$\nu_{\rm NFW}= {\cal K }_{\rm NFW}/\sigma_{\rm shape}$.
The expected $S/N$ is shown in Fig.~\ref{fig2.ps} as a
function of the concentration parameter. We find that, in the cases of
lower $\Delta$, the $S/N$ is sensitive to the concentration parameter
(the same argument but for the case of $M_{\rm vir}$ was made by
\citet{KingMead2011}). This can be explained by the fact that the
lensing $S/N$ mostly comes from the mass at the inner region of the halo
as was shown in Fig.~\ref{fig1.ps}. The change in the
concentration parameter results in the change in the mass of the inner
region, and thus results in the change in the $S/N$. On the other hand,
in the cases of higher $\Delta$,  the spherical overdensity mass
effectively defines the mass of inner region; therefore, the change in
the concentration parameter does not affect the $S/N$, though it alters
the virial mass. Therefore we can argue that the peak $S/N$ is not a
good virial mass indicator but more tightly correlates with the inner
mass such as $M_{1000}$. The result appears to be consistent with the
finding of \citet{okabe10}, who argued that weak lensing mass
measurement errors are smaller for larger overdensities of
$\Delta=500-2000$ than the case for the virial overdensity.

%%%%%%%%%%%%%%%%%%%%%%%%%%%%%%%%%%%%%%%%%%%%%%%%%%%%%%%%%%%
%%%%%%%%%%%%%%%%%%%%%%%%%%%%%%%%%%%%%%%%%%%%%%%%%%%%%%%%%%%
\section{Numerical simulations}
\label{sec:simulation}
%%%%%%%%%%%%%%%%%%%%%%%%%%%%%%%%%%%%%%%%%%%%%%%%%%%%%%%%%%%
%%%%%%%%%%%%%%%%%%%%%%%%%%%%%%%%%%%%%%%%%%%%%%%%%%%%%%%%%%%

%%%%%%%%%%%%%%%%%%%%%%%%%%%%%%%%%%%%%%%%%%%%%%%%%%%%%%%%%%
\subsection{Gravitational lensing ray-tracing simulations}
%%%%%%%%%%%%%%%%%%%%%%%%%%%%%%%%%%%%%%%%%%%%%%%%%%%%%%%%%%%

We use a large set of ray-tracing simulations that are detailed in
\citet{sato09} and \citet{OguriHamana2011}. It is based on the
ray-tracing technique developed in
\citet{HamanaMellier2001}\footnote{The ray-tracing simulation codes are 
publicly available at {\tt
    http://th.nao.ac.jp/MEMBER/hamanatk/RAYTRIX/index.html}.}. 
In what follows we describe only aspects directly relevant to this study.

The ray-tracing simulations are based on $200$ realizations of
$N$-body simulations with the box sizes of 240$h^{-1}$~Mpc (particle
mass $m_p=5.4\times 10^{10}h^{-1} M_\odot$). We use the standard
multiple lens plane algorithm to simulate gravitational lensing by
intervening matter. In this study, we consider a single source
redshift of $z_s=1$, which is a typical mean source redshift of weak
lensing analysis. Note that if redshift information on the source
galaxies, e.g. from photometric redshift, is available, one may employ
the topographic technique which improves the capability of weak lensing
cluster identification \citep{hennawi05,DietrichHartlap2010}.
Instead of using 1000 ray-tracing realizations
generated by \citet{sato09}, we regenerated 200 independent
realizations so that the same halo does not appear in different
realizations. 

Weak lensing mass maps, ${\cal K}$, are generated from lensing shear
data by applying the operation eq.~(\ref{eq:shear2kap}) on grid points
of $2048\times 2048$ with the grid spacing of 0.15~arcmin.  
We generated two types of mass maps, one is maps without the galaxy
shape noise, and the other is maps with the galaxy shape noise.
For galaxy shape noise, random ellipticities drawn from the truncated
two-dimensional Gaussian (see below) are added to shear data:
%%%%%%%%%%%
\begin{equation}
P(|e|) 
=\frac{1}{\pi \sigma_e^2 (1-e^{-1/\sigma_e^2})}
\exp\left(-\frac{e^2}{\sigma_e^2}\right),
\end{equation}
%%%%%%%%%%%
with $\sigma_e=0.4$ and assumed number density of source galaxies of
30~arcmin$^{-2}$. Peaks are identified as pixels that have higher
values of $\cal K$ than 8 surrounding pixels. Here it should be
noted that in the case of the noise added maps, especially for the
case of the PEX0 filter, crowds of high peaks caused by the galaxy
shape noise tends to form at massive cluster regions. To exclude
duplicated detections in each cluster, we apply friends-of-friends
(FOF) algorithm with the linking length of 1~arcmin \citep{maturi05},
and group together duplicated peaks as a single peak. 
This procedure generates a catalogue of lensing peaks for each map. 

%%%%%%%%%%%%%%%%%%%%%%%%%%%%%%%%%%%%%%%%%%%%%%%%%%%%%%%%%%
\subsection{Dark matter halo catalogue}
%%%%%%%%%%%%%%%%%%%%%%%%%%%%%%%%%%%%%%%%%%%%%%%%%%%%%%%%%%%

For each $N$-body output, we identify dark matter haloes using the
standard FOF algorithm with the linking parameter of $b=0.2$ and
derive a halo mass $M_{\rm FOF}$ for each halo. We define the halo
position by the potential minimum of haloes where the gravitational
potential is computed from only FOF member particles. In addition to
$M_{\rm FOF}$, we compute the spherical overdensity mass with average
overdensities of $\Delta=500$ and 1000, which are denoted by $M_{500}$
and $M_{1000}$, respectively. 

To evaluate the halo shape, we compute the inertia tensor of the mass
distribution:
\begin{eqnarray}
\label{Iij}
I_{ij}&=&\int d\bm{x}^3 (x_i-\bar{x}_i) (x_j-\bar{x}_j) \rho(\bm{x})\nonumber\\
&=& m_p \sum_{k=1}^{N_p} (x_{i,k}-\bar{x}_i) (x_{j,k}-\bar{x}_j),
\end{eqnarray}
where $i,j$ run from 1 to 3, $N_p$ is the number of FOF member
particles of haloes and $\bar{x}_i$ denotes the cluster centre defined
by the potential minimum. We use the trace of the inertia tensor to
estimate the concentration of the halo mass distribution,
\begin{equation}
\label{I}
I\equiv {1\over 3} (I_{11}+I_{22}+I_{33}).
\end{equation}
This is compared with the expected value for the uniform overdens sphere:
\begin{equation}
\label{Ith}
I_{\rm TH}= {1 \over 5} M_{\rm vir} r_{\rm vir}^2.
\end{equation}
To evaluate the triaxility of halo shape, we compute the eigenvectors
from the inertia tensor and convert them
to the axial ratios $a/c$ and $a/b$ ($a\le b\le c$).
We also compute the angle between the line of sight and the major-axis
direction of the halo which we denote as $\theta_z$.

Halo catalogues on the light cone are generated by stacking the
simulation outputs in the same manner as in the ray-tracing
experiments. In summary, each halo in the mock catalogues has data on
the mass, redshift, the angular position on the weak lensing mass map  
and the parameters for the halo shape $I$, $a/c$ and $\theta_z$.
In addition, for each halo, we compute the {\it NFW-corresponding}
peak $S/N$ using eq.~(\ref{eq:kappa_nfw}) adopting the
mass-concentration relationship eq.~(\ref{eq:c-m}) for three mass
estimates, $M_{\rm FOF}$, $M_{500}$ and $M_{1000}$, which we denote
$\nu_{\rm FOF}$, $\nu_{500}$ and $\nu_{1000}$ respectively. Also we
compute the virial radius defined by eq.~(\ref{eq:mvir}), and 
$I_{\rm TH}$ defined by eq.~(\ref{Ith}), adopting $M_{\rm FOF}$ as the
virial mass. 

%%%%%%%%%%%%%%%%%%%%%%%%%%%%%%%%%%%%%%%%%%%%%%%%%%%%%%%%%%
\section{Result}
\label{sec:results}
%%%%%%%%%%%%%%%%%%%%%%%%%%%%%%%%%%%%%%%%%%%%%%%%%%%%%%%%%%%

%%%%%%%%%%%%%%%%%%%%%%%%%%%%%%%%%%%%%%%%%%%%%%%%%%%%%%%%%%
\subsection{Halo-peak relationship}
\label{sec:halo-peak}
%%%%%%%%%%%%%%%%%%%%%%%%%%%%%%%%%%%%%%%%%%%%%%%%%%%%%%%%%%%

%
%%%%% Fig-3
%
\begin{figure}
\begin{center}
 \includegraphics[width=85mm]{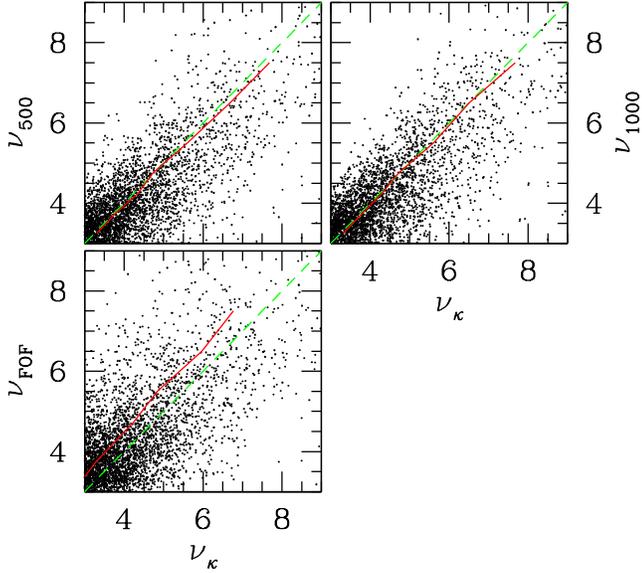}
\end{center}
\caption{Peak values in the noise-free $\cal K$ map, $\nu_{\kappa}$, is
compared with the NFW-corresponding peak $S/N$ computed from $M_{\rm FOF}$
({\it bottom left}), $M_{500}$ ({\it top left}) and $M_{1000}$
({\it top right}).  The mass maps generated from the Gaussian filter
are used. Each point shows the relation for each halo, and the solid
line shows the mean of them. Here haloes with $M_{\rm FOF}>1\times
10^{14}h^{-1} M_\odot$ at $0.1<z<0.4$ are considered.
\label{fig3.ps}}
\end{figure}

We start with matching halo catalogues with weak lensing peaks. To do
so, for each halo, the highest peak within the virial radius of the
halo (or within 3~arcmin if the virial radius is larger than 3~arcmin)
is searched for and is considered as the corresponding peak. 
We denote the peak heights by $\nu_{\kappa}$. 

First we examine the correlation between the $\nu_{\kappa}$ from the
noise-free $\cal K$ map and the NFW-corresponding peaks evaluated from
the three definitions of the halo mass, $M_{\rm  FOF}$, $M_{500}$ and
$M_{1000}$.  The aim of this analysis is to check the finding in
Section~\ref{sec:lensing-theory} that the mass definition taking into
account only the inner region  (such as $M_{1000}$) better correlates
with the peak $S/N$ than masses including outer regions (such as
$M_{200}$ or $M_{\rm vir}$). Fig.~\ref{fig3.ps}
compares $\nu_{\kappa}$ with three NFW-corresponding peaks $\nu_{\rm
  FOF}$, $\nu_{500}$ and $\nu_{1000}$, where haloes with $M_{\rm
  FOF}>1\times 10^{14}h^{-1} M_\odot$ at $0.1<z<0.4$ are considered.
To quantify the correlation, we evaluate the mean and rms scatter of
$\delta_\nu \equiv (\nu_\kappa -\nu_i)/\nu_i$ among peaks with
$\nu_i>4$, and we find (mean, RMS)$=(-0.11,0.25)$, $(0.017,0.20)$ and
$(0.0098, 0.19)$ for $i={\rm FOF}$, 500 and 1000, respectively.
Thus we find that $\nu_{1000}$ best correlates with $\nu_{\kappa}$, as
expected. Based on this result, in what follows, we take $M_{1000}$ as
the mass indicator of the haloes. Note that for the NFW halo with 
$c_{\rm vir}=4$, $M_{1000}/M_{\rm vir}\simeq 0.35$, and 
$M_{1000}/M_{\rm vir}$ is larger (smaller) for a halo with higher
(lower) concentration. We also find that $\nu_{\rm FOF}$-$\nu_{\kappa}$ 
relationships not only have larger scatters but also show some offset
in the mean relationship. One of the reasons may be the mismatch
between $M_{\rm FOF}$ computed in $N$-body simulations and 
$M_{\rm vir}$ defined by overdensity 
\citep[see also][]{OguriHamana2011}, presumably caused by 
the presence of outer substructures that contribute to the FOF mass.
In the analysis above, the mass maps generated from the Gaussian
filter are used though the results are found not to be affected by
choice of the filter. 

%
%%%%% Fig-4
%
\begin{figure}
\begin{center}
 \includegraphics[width=82mm]{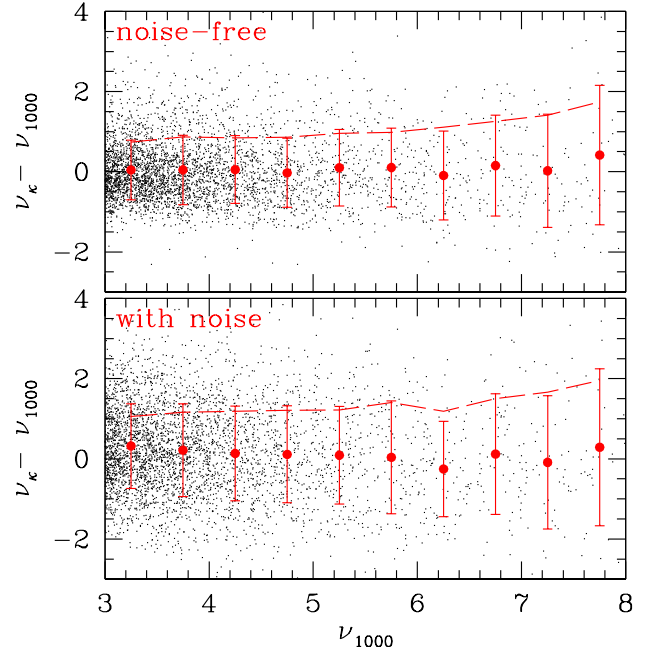}
\end{center}
\caption{Relationship between $\nu_{1000}$ and $\nu_{\kappa}$ as a function
  of $\nu_{1000}$. The dots show the relation for each halo and the
  filled circles with error bars show the mean and the rms scatter, and
  the dashed curve show the value of the rms scatter. The $\cal K$ maps
  generated from the Gaussian filter are used, and only the haloes
  with $M_{1000}>3\times 10^{13}h^{-1} M_{\odot}$ at $0.1<z<0.4$ are
  considered. The top and bottom panels show results without and with
  the shape noise, respectively.
\label{fig4.ps}}
\end{figure}

Next we examine the influence of the following three effects on the
peak $S/N$; (1) the galaxy shape noise, i.e., the noise due to the
intrinsic shapes of source galaxies used for weak lensing analysis, 
(2) the projection of structures along the line of sight of clusters,
which we call {\it the cosmic noise}, and (3) the diversity of dark
matter distributions of haloes, which we call {\it the halo shape
  effect}. To do so, the 
mass maps generated from the Gaussian filter are used though the
results are found not to be strongly dependent on the choice of the
filter. With our choice of parameters for the galaxy shape noise
(e.g., $\sigma_e=0.4$ and $n_g=30$~arcmin$^{-2}$), the rms of the
galaxy shape noise is $\sigma_{\rm shape}=0.02$, which should act as
the additive noise in $\nu_{\kappa}$. The analytic way to estimate the
cosmic noise was developed by \citet{Hoekstra2001} under the
assumption that the halo and the large-scale structures are
uncorrelated, and for the cosmological model adopted in this paper and
the Gaussian window function, its rms is estimated to be $\sigma_{\rm
  cosmic}=0.01$, which should again act as the additive noise.
The halo shape effect should, in a crude approximation, act as the
multiplicative noise in the peak $S/N$\footnote{Approximately,
  deviations of halo shapes from a fiducial spherical model can be
written as $\rho \simeq (1 + \delta)  \rho_{\rm sph}$
where $\delta$ describes deviations from the fiducial
spherical mass distribution denoted by by $\rho_{\rm sph}$.
The lensing convergence, $\kappa$, is the line-of-sight projection of
the halo mass distribution weighted by the lensing efficiency, and thus
can be approximated as $\kappa \simeq (1 + \delta) \kappa_{\rm sph}$, 
indicating that the effect of halo shapes is multiplicative.}. 
We estimate its amplitude using
the ray-tracing simulation data. We evaluate scatters in
$\nu_{\kappa}-\nu_{1000}$ relations as a function of $\nu_{1000}$, and
results are plotted in  Fig.~\ref{fig4.ps}. 

The result for the noise-free $\cal K$ map case shown in the top panel
of Fig.~\ref{fig4.ps} indicates that the rms increase
with $\nu_{1000}$. Since only the cosmic noise and halo shape effect
are playing in this shape noise free case, and the cosmic noise should
be independent of the $S/N$, the dependence of the rms on
$\nu_{1000}$ is most likely originated from the halo shape effect. 
We estimate from the rms measured from $\nu_{\kappa}-\nu_{1000}$
relationship and the expectation value of $\sigma_{\rm cosmic}=0.01$
(thus $\sigma_{\rm cosmic}/\sigma_{\rm shape}=0.5$) that the rms of
the halo shape effect scales approximately $\sigma_{\rm halo}\sim
(0.1-0.2) \times {\cal  K}_{NFW}$ (i.e., $\sigma_{\rm
  halo}/\sigma_{\rm shape}\sim (0.1-0.2)\times\nu$). Thus for very massive
haloes the halo shape effects can be larger than the other noises
(the qualitatively same argument was made by \cite{TangFan2005}). 
However, it should be noted that for a very shallow survey the above
argument may not be the case as $\sigma_{\rm shape}$ scales with the
source galaxy number density as $\sigma_{\rm shape} \propto n_g^{-0.5}$.
In the noise added case plotted in the bottom panel of
Fig.~\ref{fig4.ps}, the mean value is slightly greater
than zero, quantitatively $\langle \nu_{\kappa}-\nu_{1000}\rangle \sim
0.25$ thus $\sim 5$ per cent for the $\nu=5$ peak. This is because we
consider the peaks that are affected by the shape noise in an
asymmetric way as discussed in \citet{hamana04}. 

%
%%%%% Fig-5
%
\begin{figure}
\begin{center}
 \includegraphics[width=82mm]{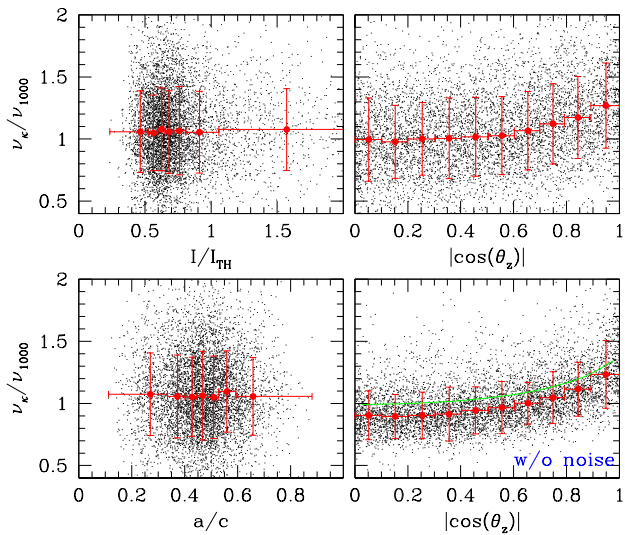}
\end{center}
\caption{Relationship between $\nu_{1000}$ and $\nu_{\kappa}$ as a
  function of $I/I_{\rm TH}$ ({\it top left}), $a/c$ ({\it bottom
    left}) and $|\cos(\theta_z)|$ ({\it right panels}).
Here haloes with $M_{1000}>3\times 10^{13}h^{-1} M_{\odot}$ at $0.1<z<0.4$
are considered.
 The dots show the relationship for each halo and filled circles show
 mean, vertical error bars show the rms scatter, and the horizontal
 error bars show the range where sample is taken. The $\cal K$ maps
 with the shape noise generated from the Gaussian filter are used
 except for the bottom-right panel where the noise-free $\cal K$
 maps are used. The green curve in the bottom-right panel shows
 the theoretical prediction based on the triaxial halo model.
\label{fig5.ps}}
\end{figure}

We now discuss details of the halo shape effect. In
Fig.~\ref{fig5.ps}, the fractional difference 
between $\nu_{1000}$ and $\nu_{\kappa}$ measured from $\cal K$ map
(with the galaxy shape noise added) is shown as a function of the halo
concentration $I/I_{\rm TH}$ (top-left panel), the axis ratio $a/c$
(bottom-left panel) and the halo orientation with respect to the
line-of-sight direction $|\cos(\theta_z)|$ (right panels). Dots
indicate values for each halo, where we consider haloes with
$M_{1000}>3\times 10^{13}h^{-1} M_\odot$ at $0.1<z<0.4$ with the mean $S/N$ of
this halo sample being $\langle \nu_{1000} \rangle=3.7$. Filled 
circles show average values. We find that there exists (1) a clear
correlation between the halo orientation and the peak height
deviations from the NFW model prediction, whereas (2) no correlation
for the halo shape parameters ($I/I_{\rm TH}$ and $a/c$). The results
indicate that the intrinsic halo shape (concentration and axis ratio)
does not cause a systematic bias in the peak heights {\it as long as
one employs an appropriate definition of the halo mass } such as
$M_{1000}$,  
but just contribute to the scatter. However the halo orientation does
cause the systematic bias because the line-of-sight projected mass at
the inner region depends strongly on the halo orientation \citep[see
also][]{oguri05,gavazzi05}. In the bottom-right panel of
Fig.~\ref{fig5.ps}, the theoretical prediction
based on the triaxial halo model \citep{JingSuto2002,ob09} is also
shown (see Section~\ref{sec:halo-theory}). We find that the triaxial
model nicely reproduces the orientation dependence of the peak heights
found in ray-tracing simulations, except for the small offset which is
originated from an approximation involved in the triaxial haloes 
\citep[see][]{oguri05}.

%%%%%%%%%%%%%%%%%%%%%%%%%%%%%%%%%%%%%%%%%%%%%%%%%%%%%%%%%%
\subsection{Selection bias in weak lensing selected clusters}
%%%%%%%%%%%%%%%%%%%%%%%%%%%%%%%%%%%%%%%%%%%%%%%%%%%%%%%%%%%

%
%%%%% Fig-6
%
\begin{figure}
\begin{center}
 \includegraphics[width=82mm]{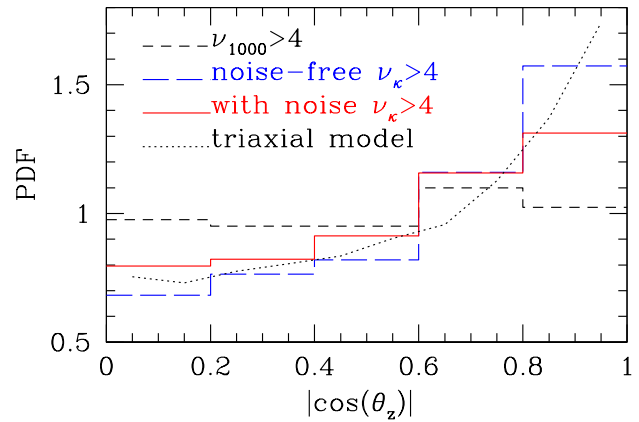}
\end{center}
\caption{Probability distribution functions of the halo orientation
  with respect to the line-of-sight direction, $|\cos(\theta_z)|$, 
  for weak lensing selected clusters. We adopt the peak height
  threshold value of $\nu=4$. The dashed histogram is for the
  NFW-corresponding peaks, i.e., peak heights evaluated by the spherical
  NFW model via spherical overdensity masses, the long-dashed
  histogram for the peak heights measured from the noise-free $\cal K$
  map, and the solid histogram for the peak heights measured from the
  noise added $\cal K$ map. The dotted curve shows the analytic model
  expectation based on the triaxial halo model. 
\label{fig6.ps}}
\end{figure}

%
%%%%% Fig-7
%
\begin{figure}
\begin{center}
\includegraphics[width=82mm]{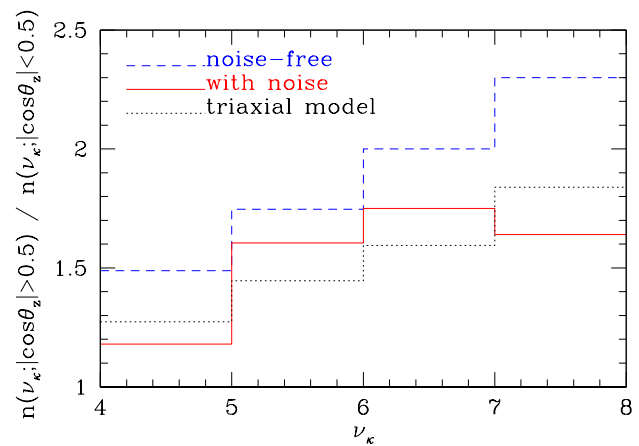}
\end{center}
\caption{The halo orientation bias as a function of the peak $S/N$
  threshold. We quantify the bias by the ratio of the numbers of
  haloes with the major axis aligned to the line-of-sight direction
  ($|\cos \theta_z |>0.5$) to those with anti-aligned ($|\cos \theta_z
  |<0.5$). The dashed histogram is for the peak heights measured from
  the noise-free $\cal K$ map, and the solid histogram for the peak
  heights measured from the noise added $\cal K$ map. The dotted
  histogram shows the analytic model prediction based on the triaxial 
  halo model. 
\label{fig7.ps}}
\end{figure}

Given the strong dependence of the peak heights on the halo
orientation, we shall now investigate its impact on weak lensing
selected cluster catalogues. 
In Fig.~\ref{fig6.ps}, we show the
probability distribution function (PDF) of the halo orientation
  with respect to the line-of-sight direction, 
$|\cos(\theta_z)|$, for haloes with peak heights above the threshold 
value $\nu=4$. Here we include only haloes at $0.1<z<0.4$ for
simplicity. Different histograms are for different causes;
NFW-corresponding peaks 
(dashed), peak heights measured from the noise-free $\cal K$ map
(long-dashed), and the noise added map (solid). In addition, the
same PDF computed with analytic triaxial halo  model (see
Section~\ref{sec:halo-theory}) is also plotted by the dotted line,
which should be compared with the noise-free case because the galaxy
shape noise is not added in the triaxial model calculation. The PDF
for the NFW-corresponding peaks ($\nu_{1000}$) is flat. This is exactly
expected because the NFW-corresponding peak heights are computed from
$M_{1000}$ assuming the spherical NFW profile, and hence they should not
depend on the orientations of haloes.  
On the other hand, the PDFs for the measured
peaks are significantly skewed such that the major axes of haloes are
preferentially aligned with the line-of-sight direction. To quantify
the bias, we introduce an estimator defined by the ratio between the
numbers of haloes with the major axis aligned to the 
line-of-sight direction ($|\cos \theta_z |>0.5$) and anti-aligned 
($|\cos \theta_z |<0.5$), i.e., $b_z \equiv n(|\cos \theta_z
|>0.5)/n(|\cos \theta_z |<0.5)$. For the cases shown in 
Fig.~\ref{fig6.ps}, the bias are 
$b_z =1.7$ and 1.4 for the noise-free and noise added cases,
respectively. For the case of the NFW-corresponding peaks we have $b_z
= 1.07$, suggesting that the error in the above estimation is less
than 10 per cent. 

In addition, we examine the bias as a function of the peak heights
using the estimator, $b_z(\nu_\kappa) \equiv n(\nu_\kappa;|\cos \theta_z
|>0.5)/n(\nu_\kappa;|\cos \theta_z |<0.5)$. The results are shown in
Fig.~ \ref{fig7.ps}. While the statistic is not very
good especially for higher $S/N$ bins because of a limited 
number of peaks, one can clearly see a trend that the orientation bias
is larger for higher $S/N$. This is presumably because the halo shape
effect is multiplicative to the $S/N$. Although the results are not
accurate enough to make precise predictions, we find that the bias is
$\sim 1.5$ for high $S/N$ peaks of $\nu_\kappa\ga5$. and $\sim 1.2$
for $\nu_\kappa\sim4$. Therefore the orientation bias is one of the
most significant selection bias in weak lensing selected cluster
catalogues. Its impact on cluster related sciences should be taken
into consideration. 
The results shown in Figs.~\ref{fig5.ps}, 
\ref{fig6.ps}, and
\ref{fig7.ps} indicates that the analytic triaxial model
predictions are in very good agreement with the trends found in
ray-tracing simulations. This confirms that the orientation bias comes
from the intrinsic triaxial shapes of haloes, rather than the effect of
surrounding matter around clusters, because such correlated matter is
not included in our analytic triaxial model calculations. 

We note that the strong halo orientation bias was also found in
previous strong lensing studies as well. \citet{hennawi07} used a
large set of ray-tracing simulations to show that major axes of clusters
producing giant arcs are preferentially aligned with the line-of-sight 
direction, with the level of the bias similar to what found in this
paper. \citet{ob09} reported that clusters having larger Einstein radii
have larger orientation bias, based on the triaixl halo model. Similar
result for strong lensing clusters was obtained also by
\citet{meneghetti10}. Our result indicates that similar orientation
bias exists for weak lensing selected clusters as well.

%%%%%%%%%%%%%%%%%%%%%%%%%%%%%%%%%%%%%%%%%%%%%%%%%%%%%%%%%%
\subsection{Completeness}
%%%%%%%%%%%%%%%%%%%%%%%%%%%%%%%%%%%%%%%%%%%%%%%%%%%%%%%%%%%

%
%%%%% Fig-8
%
\begin{figure}
\begin{center}
 \includegraphics[width=84mm]{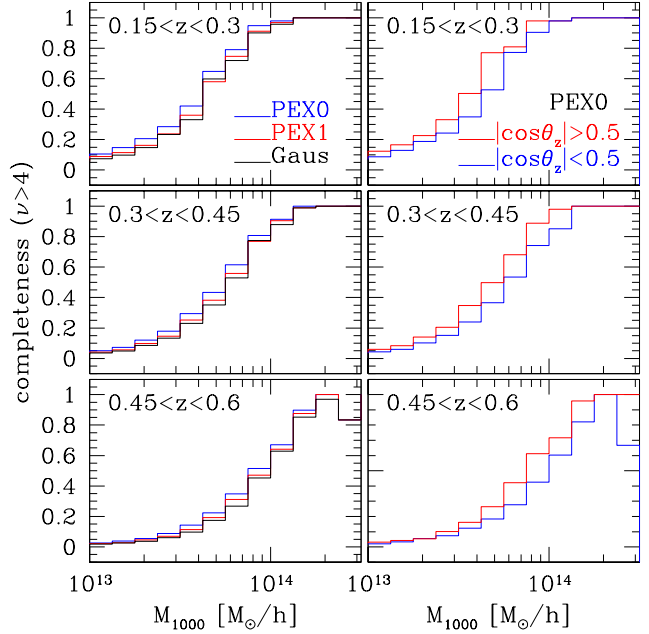}
\end{center}
\caption{Completeness as a function of the halo mass. From top to
  bottom panels, we change the redshift range from lower to higher
  mean redshifts. {\it Left panels:} Different histograms show results
  for different filters. Blue, red, and black histograms are  for the
  PEX0, PSX1 and Gaussian filters, respectively. 
  {\it Right panels:} The red histogram is for the the haloes with
  $|\cos(\theta_z)|>0.5$, i.e., haloes with major axes aligned with
  the line-of-sight direction,  whereas the blue histogram is for
  anti-aligned haloes.
\label{fig8.ps}}
\end{figure}

To see the impact of the halo orientation bias, we examine the 
{\it completeness} which we define by the fraction of haloes
identified as high peaks above a given threshold ($\nu_{min}$) in the
$\cal K$ map relative to all the haloes, namely,  $n(M,z;
\nu_\kappa>\nu_{min})/n(M,z)$.  The choice of the threshold is
arbitrary, usually it is chosen considering the trade-off between the 
sample size and the fraction of the false positives (see 
Section~\ref{sec:peakcounts}). Here we adopt $\nu_{min}=4$.
The completeness for different filters are shown in left panels of
Fig.~\ref{fig8.ps}. We find that the PEX0
filter is the best in terms of the completeness, although the
difference between filters is small, 10 per cent at most.
Right panels of Fig.~\ref{fig8.ps} compares 
the completeness for haloes with $|\cos \theta_z |>0.5$ (red
histogram) with the others (blue), where the $\cal K$ maps with PEX0
filter are used. We find that the effect of the halo orientation bias
is visible for the intermediate mass haloes for which 
$\nu_{\rm NFW}\sim\nu_{\rm min}$. The reason is that for the very
massive/small haloes, the $S/N$ differs very much from $\nu_{\rm
  min}$, meaning that the spread of $\nu$ produced by the halo shape
effect does not affect the detectability of those haloes. 
Thus the completeness of haloes with $\nu_{\rm NFW}\sim \nu_{min}$ are mostly affected 
by the halo orientation bias. The difference of the amplitude for the
completeness is about 20 per cent for those haloes. 

%%%%%%%%%%%%%%%%%%%%%%%%%%%%%%%%%%%%%%%%%%%%%%%%%%%%%%%%%%
\subsection{Peak counts and purity}
\label{sec:peakcounts}
%%%%%%%%%%%%%%%%%%%%%%%%%%%%%%%%%%%%%%%%%%%%%%%%%%%%%%%%%%%

%
%%%%% Fig-9
%
\begin{figure}
\begin{center}
 \includegraphics[width=82mm]{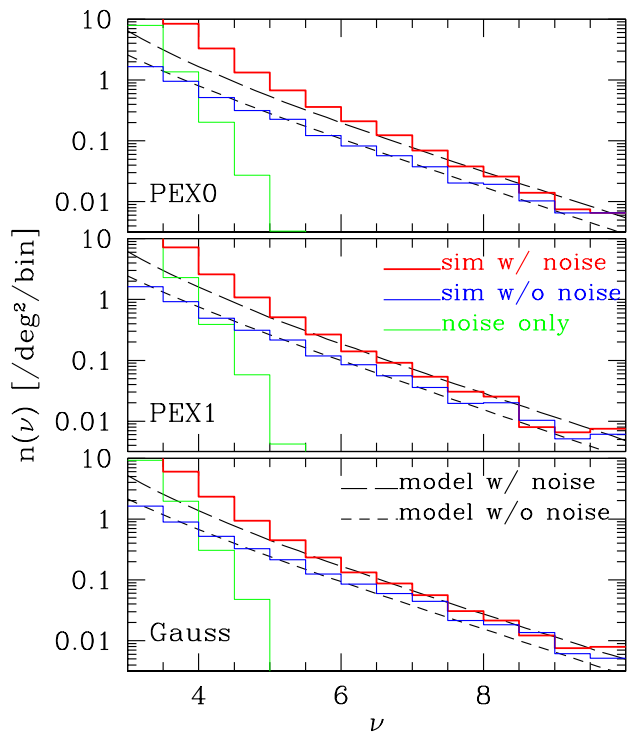}
\end{center}
\caption{Number counts of weak lensing peaks for three filters from
  top to bottom panels. The red and blue histograms are measurements
  from ray-tracing simulations with and without galaxy shape noise,
  respectively. The green histogram is the number count of peaks from
  the pure noise map. The long-dashed and dashed curves are analytical
  predictions using the spherical NFW profile, with and without shape
  noise taken into account, respectively. 
\label{fig9.ps}}
\end{figure}

%
%%%%%  Table 1
%
\begin{table}
\caption{Summary of the number density (per 1~deg$^2$)
of weak lensing mass peaks (in noise added $\cal K$ maps) 
above thresholds.}
\label{table:peakcounts}
\begin{tabular}{lccc}
\hline
filter & $N_{\rm peak}(\nu_\kappa>4)$  & $N_{\rm peak}(\nu_\kappa>5)$  &
$N_{\rm peak}(\nu_\kappa>6)$  \\ 
\hline
PEX0  & 3.3 & 0.70 & 0.22\\
PEX1  & 2.7 & 0.59 & 0.19 \\
Gauss & 2.2 & 0.53 & 0.19 \\
\hline
\end{tabular}
\end{table}

So far we discussed connections between the weak lensing properties of
clusters and dark matter halo properties. Here, we study the subject
from a different viewpoint, i.e., we examine properties of the weak
lensing peak sample in terms of {\it peak counts} and {\it purity},
paying special attention to the comparison among three filters.

First we examine peak counts and compare with the analytic model.
In Fig.~\ref{fig9.ps}, peak counts measured
from three $\cal K$ maps (the galaxy noise free, noise added, and the
pure noise cases) are presented for three filters. 
The number densities of peaks (in noise added $\cal K$ maps) above the given
thresholds are summarized in Table~\ref{table:peakcounts}. We find that
the PEX0 filter yields the largest peak counts, specifically about 20
per cent larger than  the PEX1 filter, and even larger difference from
the Gaussian filter. This is a natural consequence that the PEX0
filter mimics the optimal filter developed by \cite{maturi05} which
is designed for maximizing the $S/N$ of $\cal K$ from the NFW halo 
(see \cite{pace07} for the test of the capability of the optimal filter against
numerical simulations).
To compute the analytic prediction of the peak
counts, we follows the so-called halo model first developed by
\citet{KruseSchneider2000} \citep[see
also][]{bartelmannetal2001,hamana04}, in which it is assumed that 
high peaks are dominated by lensing signals from single massive haloes.
Here we adopt the spherical NFW density profile for the analytic
calculation of the number count, because we include the effect of the
halo triaxiality via the halo shape noise as described below. 
To take into account the effect of noise, we employ the approximate
approach developed by \citet{hamana04} with some modification.
Specifically, we make the following three assumptions.
(1) Very high peaks are neither removed nor generated by the noise but
their peak heights are altered by the noise. 
(2) The scatter in peak heights with respect to the corresponding NFW
peak height follows the Gaussian distribution with the standard
deviation of $\sigma_{\rm peak}$. While in \citet{hamana04} only the
galaxy shape noise was taken into accounted, here we include both the
cosmic noise and halo shape noise, in addition to the galaxy shape
noise. (3) Within a small range of the peak height ($\nu$),
the peak counts can be approximated by the exponential form
$n_{\rm peak}(\nu)=n_\ast\exp(p\nu)$, with a constant exponential
index of $p$. Under these assumptions the peak counts in the presence
of the noises are given by \citep{hamana04},
\begin{equation}
\label{napprox}
n_{\rm noisy}(\nu) \simeq \exp\left(
{{f^2 p^2} \over {2}}
\right) n_{\rm peak}(\nu),
\end{equation}
where $f=\sigma_{\rm peak}/\sigma_{\rm noise}$. To estimate $f$, we
assume the proportion of $\sigma_{\rm shape}:\sigma_{\rm
  cosmic}:\sigma_{\rm halo} = 2:1:1$. For the case of the Gaussian
filter, this corresponds to $\sigma_{\rm halo}=0.01$ which is a
reasonable approximation for peaks with $\nu\sim 5$ (see 
Section~\ref{sec:halo-peak}). We take the quadrature sum of these
three components to obtain $f\simeq 1.2$. The peak counts computed in
this method are presented in
Fig.~\ref{fig9.ps}.  
We find that for high $S/N$ of $\nu\ga5$ the improved analytic
prediction agrees well with the measurements, whereas for lower 
$S/N$ the measured counts are slightly larger than the prediction. 
A possible origin of the excess is the false signals which we examine 
below.

%
%%%%% Fig-10
%
\begin{figure}
\begin{center}
 \includegraphics[width=82mm]{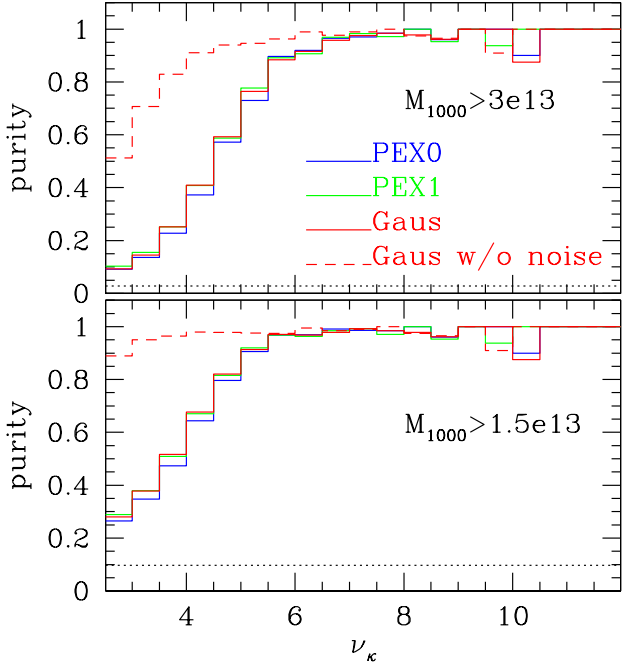}
\end{center}
\caption{The purity for the three filters are plotted by lines in
  different colors. The dashed histogram is for the noise-free case.
  Top panel is for the case where peaks are matched with haloes
  with $M_{1000}>3\times 10^{13}h^{-1} M_\odot$, whereas bottom panel is
  for $M_{1000}>1.5\times 10^{13}h^{-1} M_\odot$. The horizontal dotted
  line shows the probability that a peak is matched with a randomly
  distributed halo by chance. 
\label{fig10.ps}}
\end{figure}

Finally, we examine the {\it purity} which we here define by the
fraction of peaks which are associated with true haloes among 
all the peaks. However, in practice, the purity is not uniquely
defined quantity because of the following two ambiguities. 
One is the allowed separation between peak position and halo position.
The other is the minimum mass of haloes allowed to be associated with
peaks. We adopt the minimum halo mass of
$M_{1000}=3\times10^{13}h^{-1} M_\odot$ or $1.5\times10^{13}h^{-1} M_\odot$
based on the detectability study shown in
Fig.~\ref{fig8.ps}. Considering the virial
radii of those haloes, we fix the maximum separation at 3~arcmin.
The results are presented in Fig.~\ref{fig10.ps}.
We also plot the probability that a peak matched with a randomly
distributed halo (with the same number density as the true halo
catalogue) by chance by the horizontal dotted line, which
gives an estimate of the chance coincidence between peaks and
unrelated haloes. We find that the probability of the chance matching
is not significant for our choice of the minimum masses and the
allowed separation. The purity is not dependent on the choice of the
filter, and is very high (i.e., the false positive rate is very small)
for peaks with $\nu>6$. However the purity drops rapidly at lower
$S/N$, and it becomes about 50 per cent at $\nu\sim4$. 
We argue that this accounts for the excess in the peak counts over the
theoretical prediction found in Fig.~\ref{fig9.ps}.
In the same figure, we also plot the purity measured from the galaxy
shape noise free case. The false positives in this case may arise from
chance projections such as the line-of-sight projection of small
multiple haloes or filamentary structure. As the purity for the
noise-free case is found to be greater than 90 per cent for $\nu>4$,
we conclude that the contamination of such chance projection is not
significant. 

%%%%%%%%%%%%%%%%%%%%%%%%%%%%%%%%%%%%%%%%%%%%%%%%%%%%%%%%%%%
%%%%%%%%%%%%%%%%%%%%%%%%%%%%%%%%%%%%%%%%%%%%%%%%%%%%%%%%%%%
\section{Comparison with observations}
\label{sec:scam}
%%%%%%%%%%%%%%%%%%%%%%%%%%%%%%%%%%%%%%%%%%%%%%%%%%%%%%%%%%%
%%%%%%%%%%%%%%%%%%%%%%%%%%%%%%%%%%%%%%%%%%%%%%%%%%%%%%%%%%%
 
%%%%% Fig-11
%
\begin{figure}
\begin{center}
 \includegraphics[width=82mm]{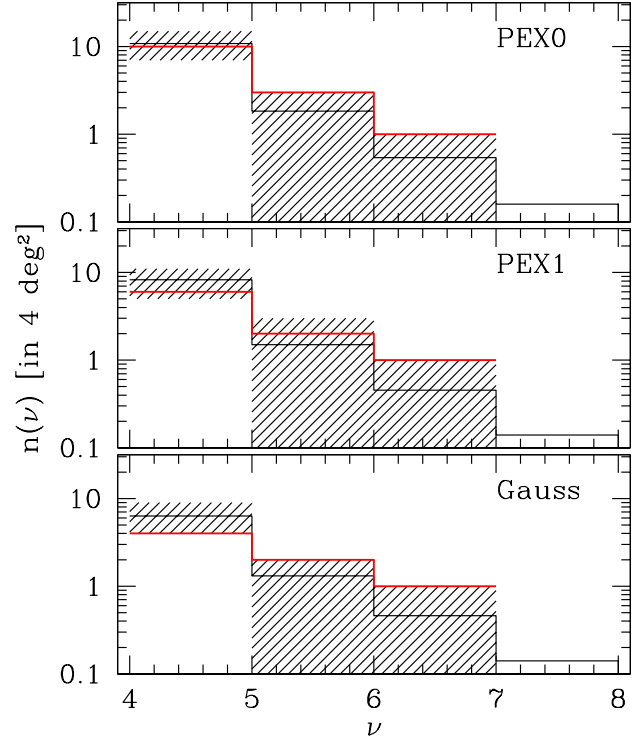}
\end{center}
\caption{Red histogram shows the number count of peaks measured from
  real weak lensing $\cal K$ maps generated from 4~deg$^2$ of
  Suprime-Cam data (see Appendix~\ref{sec:subaru} for
  details). Results from mock simulation of 4~deg$^2$ survey are
  shown by black histogram (the average among 200 realizations) with
  shading (the range enclosing 68 per cent). Note that on the highest
  bin, peak counts are zero in most of the realizations and non-zero
  counts are measured in a small fraction of realizations, therefore
  the average is non-zero but the 68 per cent enclosing region is not
  defined.
\label{fig11.ps}}
\end{figure}

It is worth checking the results of our mock simulation data against
real observational results. In Fig.~\ref{fig11.ps}, we
compare the peak counts measured from real weak lensing $\cal K$ maps
generated from 4~deg$^2$ of Subaru/Suprime-Cam data with results from
the mock
simulations. A description of the Suprime-Cam data and data analysis
is given in Appendix~\ref{sec:subaru}. In short, the Suprime-Cam data
consist of four fields with fairly uniform data quality (with respect
to the depth and seeing condition) with the number density of galaxies
used for weak lensing analysis is $n_g\simeq 27$~arcmin$^{-2}$ and the
rms ellipticity of $\sigma_{\rm shape}=0.4$ which are in a good
agreement with the values adopted in the mock simulations. The mean
source redshift is not known as there is no large
spectroscopic/photometric redshift galaxy catalogue reaching the depth
of the  data ($i'\sim25.5$ AB mag), but the value of $\langle z_s
\rangle=1$ assumed in mock simulations is reasonable \citep[see
also][]{oguri12}. We generated $\cal K$ maps with the three filters
under consideration, and we searched for peaks in the $\cal K$ maps. 
The peaks located within 1 arcmin from the field boundary were
discarded as the regions are likely affected by the partial lack of
data. The total area used for peak finding is 4~deg$^2$. 
We detect 14, 9 and 7 peaks with $\nu>4$ for the PEX0, PEX1 and
Gaussian filter, respectively. To measure the peak counts from mock
simulations under a similar survey condition, we extracted a
contiguous 4~deg$^2$ region from each mock weak lensing
realization, and measured peak counts. We computed the average counts
and the range enclosing 68 per cent of 200 realizations and show in
Fig.~\ref{fig11.ps}.  
We find that (i) peak counts from real data are in a reasonable
agreement with the expectation from mock data, and (ii) its dependence
on the filter is similar to that expected from the mock data.
Of course this comparison tests only a limited aspect of the mock
simulation, yet the agreements can be regarded as a piece of evidence
that our numerical simulations based on the standard cosmological
model and realistic noise parameters produce reasonably realistic mock
data.

%%%%%%%%%%%%%%%%%%%%%%%%%%%%%%%%%%%%%%%%%%%%%%%%%%%%%%%%%%%
%%%%%%%%%%%%%%%%%%%%%%%%%%%%%%%%%%%%%%%%%%%%%%%%%%%%%%%%%%%
\section{Summary and discussions}
\label{sec:summary}
%%%%%%%%%%%%%%%%%%%%%%%%%%%%%%%%%%%%%%%%%%%%%%%%%%%%%%%%%%%
%%%%%%%%%%%%%%%%%%%%%%%%%%%%%%%%%%%%%%%%%%%%%%%%%%%%%%%%%%%
  
We have investigated scatter and bias in weak lensing selected
clusters, employing both the analytic model description of dark matter
haloes and the numerical mock data of weak lensing cluster surveys 
generated with gravitational lensing ray-tracing through a large set
of $N$-body simulations. We have paid special attention to the effects of
diversity in the dark matter distribution of haloes. Our major
findings are summarized as follows. 

\def\theenumi{(\arabic{enumi})}
\begin{enumerate}
\item We have examined the relationship between the peaks measured
  from the noise-free $\cal K$ map and the expected peak values
  computed assuming the spherical NFW profile with three cluster mass 
  definitions, $M_{\rm FOF}$, $M_{500}$ and $M_{1000}$. We have found
  that the expected peak value computed with $M_{1000}$ best
  correlates with the measured peak heights. This confirms the finding
  in Section~\ref{sec:lensing-theory} that the lensing $S/N$ computed
  with the spherical overdensity mass with higher mean interior
  density is less affected by variation in halo concentrations than
  those with lower interior density. An implication to observational
  studies is that the peak $S/N$ is not a good virial mass indicator
  but is more tightly related to the inner mass such as $M_{1000}$. 
\item We have examined the influence of the diversity of the halo
  shape on the peak $S/N$ values by comparing the peaks measured from the
  noise-free $\cal K$ map with the NFW-corresponding peak heights
  computed with $M_{1000}$ using the spherical NFW profile. We have
  found that the rms of the halo shape effect scales approximately
  $\sigma_{\rm halo}\sim (0.1-0.2)\times {\cal  K}_{NFW}$ (i.e.,
  $\sigma_{\rm halo}/\sigma_{\rm shape}\sim (0.1-0.2)\times\nu$). We
  therefore conclude that the scatter caused by the halo shape effect
  can be larger than the ones from other noises for very massive haloes.
\item We have found a clear correlation between the halo orientation
  and the peak heights, such that haloes whose major axes are aligned
  with the line-of-sight direction tend to generate a higher peak than
  the expectation of the spherical NFW model.  Using both numerical
  and analytic approaches, we have examined the systematic bias caused
  by the orientation effect. We have evaluated the bias using the
  ratio $b_z$ between the numbers of haloes (identified with peak $S/N$
  above certain thresholds) with the major axis aligned with the
  line-of-sight direction ($|\cos \theta_z |>0.5$) and those
  anti-aligned. We found that the bias is $b_z = 1.4$ for the
  threshold of $\nu_{ min} =4$, indicating that orientation bias is
  quite strong. In addition, we have found that the bias is larger for
  higher $S/N$ peaks, because the halo shape effect is multiplicative
  to $S/N$. Thus the orientation bias is a non-negligible selection
  bias in weak lensing selected cluster catalogues. We have also
  examined the effect of the halo orientation bias on the completeness
  and found that the haloes with $\nu_{\rm NFW}\sim \nu_{min}$ are
  most affected by the halo orientation bias, and its amplitude for
  the completeness is found to be about 20 per cent for those haloes. 
  We have shown that the analytic triaxial halo model explains the
  orientation bias found in ray-tracing simulations very well, which
  confirms that the halo triaxiality is the main cause of the
  orientation bias. 
\item We have compared the capability of three filters with respect to
  the the completeness, peak counts and purity. We have found that the
  PEX0 works best, which gives about 10 per cent better completeness
  for haloes with $\nu \sim \nu_{min}$ than the other two filters and
  about $\ga 20$ per cent larger peak counts with the similar level of
  purity. 
\item We have developed a prescription to analytically compute the
  number count of weak lensing peaks which improve earlier works 
  \citep{KruseSchneider2000,bartelmannetal2001,hamana04}. 
  Specifically, we have employed an approximate way to include the
  scatters in peak heights caused by all the three sources addressed in
  this paper; the galaxy shape noise, the cosmic
  noise, and the halo shape noise. We have tested the improved model
  against the mock ray-tracing simulation data, and found that the
  analytic predictions agree well with the simulation results for
  a high $S/N$ of $\nu\ga5$. 
\item We have compared the number counts of peaks from the numerical
  mock data with those from 4~deg$^2$ of real Subaru/Suprime-cam
  data. Although the comparison is limited by small number statistics, 
  we have found that both the number counts and their smoothing filter
  dependence are in reasonable agreement between the mock data and the
  observation.
\end{enumerate}

%%%%%%%%%%%%%%%%%%%%%%%%%%%%%%%%%%%%%%%%%%%%%%%%%%%%%%%%%%%%%%
%%%%%%%%%%%%%%%%%%%%%%%%%%%%%%%%%%%%%%%%%%%%%%%%%%%%%%%%%%%%%%
\section*{Acknowledgments}
%%%%%%%%%%%%%%%%%%%%%%%%%%%%%%%%%%%%%%%%%%%%%%%%%%%%%%%%%%%%%%
%%%%%%%%%%%%%%%%%%%%%%%%%%%%%%%%%%%%%%%%%%%%%%%%%%%%%%%%%%%%%%

We thank M. Maturi, M. Takada, and N. Yoshida for useful discussions.
We thank Y. Utsumi for help with photometric calibration of
Suprime-Cam data.
M. Shirasaki and M. Sato are supported by a Grant-in-Aid for the Japan
Society for Promotion of Science (JSPS) fellows.
Numerical computations in the paper were in part carried out on the
general-purpose PC farm at Center for Computational Astrophysics, CfCA,
of National Astronomical Observatory of Japan.
This work is based in part on data collected at Subaru Telescope and
obtained from the SMOKA, which is operated by the Astronomy Data Center,
National Astronomical Observatory of Japan.
This work is supported in part by Grant-in-Aid for
Scientific Research from the JSPS Promotion of Science
(23540324, 23740161), the FIRST program "Subaru Measurements of Images
and Redshifts (SuMIRe)", and World Premier International Research Center
Initiative (WPI Initiative), MEXT, Japan. 

%%%%%%%%%%%%%%%%%%%%%%%%%%%%%%%%%%%%%%%%%%%%%%%%%%%%%%%%%%%%%%%%%%%%%%%
%  ref.tex   Time-stamp: <2012-06-23 11:38:25 hamana>
%%%%%%%%%%%%%%%%%%%%%%%%%%%%%%%%%%%%%%%%%%%%%%%%%%%%%%%%%%%%%%%%%%%%%%%

\appendix

%%%%%%%%%%%%%%%%%%%%%%%%%%%%%%%%%%%%%%%%%%%%%%%%%%%%%%%%%%%
%%%%%%%%%%%%%%%%%%%%%%%%%%%%%%%%%%%%%%%%%%%%%%%%%%%%%%%%%%%
\section{Weak lensing analysis of Suprime-Cam data}
\label{sec:subaru}
%%%%%%%%%%%%%%%%%%%%%%%%%%%%%%%%%%%%%%%%%%%%%%%%%%%%%%%%%%%
%%%%%%%%%%%%%%%%%%%%%%%%%%%%%%%%%%%%%%%%%%%%%%%%%%%%%%%%%%%
We collected $i'$-band data taken with the Subaru/Suprime-Cam
\citep{MiyazakiSCam} from the data archive {\it SMOKA}\footnote{\tt
  http://smoka.nao.ac.jp/}, under the following three conditions: data
are contiguous with at least four pointings, the exposure time for each
pointing is longer than 1800~sec, and the seeing full width at
half-maximum (FWHM) is better than
0.65 arcsec. Four data sets meet these requirements and are summarized
in Table~\ref{table:fields}. 

%
%%%%%  Table A-1
%
\begin{table}
\caption{Summary of data sets. See Figs.~\ref{fig:A1}-\ref{fig:A4} for
  the coordinate of each field}
\label{table:fields}
\begin{tabular}{lcccc}
\hline
field name & area$^{a}$ & $T_{exp}$$^{b}$ & seeing$^{c}$ &
$n_g$$^{d}$\\ 
\hline
SXDS         & 0.99 (0.77) & 6000 & 0.52 & 27.6 \\
COSMOS       & 1.85 (1.32) & 2400 & 0.53 & 26.5 \\
Lockman-hole & 0.74 (0.56) & 3600 & 0.51 & 27.3 \\
ELAIS N1     & 1.99 (1.39) & 3600 & 0.57 & 26.6 \\
\hline
\end{tabular}\\
$^{a}$ Area after masking regions affected by bright stars in unit of
deg$^2$. The numbers in parentheses are the area after
cutting the edge regions within 1~arcmin from the boundary.\\
$^{b}$ Exposure time for each pointing in the unit of seconds.\\
$^{c}$ Median of seeing full width at half-maximum in the unit of arcsec.\\
$^{d}$ Number density of galaxies used for the weak lensing analysis 
in the unit of arcmin$^{-2}$.\\
\end{table}

Each CCD data was reduced 
using the {\it SDFred}\footnote{In the process of the correction for both
  the field distortion and differential atmospheric dispersion, the
  bi-cubic resampling scheme was implemented to suppress the aliasing effect
  \citep{HamanaMiyazaki2008}.} software \citep{Yagietal2002,Ouchietal2004}.
Note that we conservatively use the data only within 15 arcmin radius
from the field centre of Suprime-Cam, because at the outside of that,
the point spread function (PSF) becomes elongated significantly which
may make the correction for the PSF inaccurate. Then mosaic stacking
was done with {\it SCAMP} \citep{SCAMP} and {\it SWarp}\footnote{{\it
    SWarp} was modified so that it can treat the bad pixel flag from
  the {\it SDFred} software properly.} \citep{SWarp}. Object
detections were performed with {\it SExtractor} \citep{SExtractor} 
and {\it hfindpeaks} of {\it IMCAT} software \citep{KSB95}, and two
catalogues were merged by matching positions with a tolerance
of 1~arcsec. 

For weak lensing measurements, we follow the so-called KSB method
described in \cite{KSB95}, \cite{LuppinoKaiser1997}, and
\cite{Hoekstraetal1998}. Stars are selected in a standard way by
identifying the appropriate branch in the magnitude half-light radius
($r_h$) plane, along with the detection significance cut $S/N>10$. 
The number density of stars is found to be $\sim 1$~arcmin$^{-2}$ for
the four fields. We only use galaxies met the following three
conditions, (i) the detection significance of $S/N>3$ and 
$nu>10$ where $nu$ is an estimate of the peak significance given by
$hfindpeaks$, (ii) $r_h$ is larger than the stellar branch, and
(iii) the AB magnitude is in the range of $22 <i'<25.5$ (where
MAG\_AUTO given by the {\it SExtractor} is used for the magnitude).
The number density of resulting galaxy catalogue is quite uniform not 
only over each field but also among four fields (see Table \ref{table:fields}).
We measure the shapes of objects by {\it getshapes} of {\it IMCAT}, and
correct for the PSF by the KSB method. The rms of the galaxy
ellipticities after the PSF correction is found to be 0.4 for all the
four fields. 

%
%%%%%  Table A-2
%
\begin{table}
\caption{Summary of high peaks.}
\label{table:peaks}
\begin{tabular}{lccccc}
\hline
field-ID & R.A. & Dec. & \multicolumn{3}{c}{$\nu$ (Peak-$S/N$)}\\ 
{} & [deg] & [deg] & PEX0 & PEX1 & GAUSS\\ 
\hline
SXDS-1     &  34.0185 & $-5.0755$  & 4.1 & 3.8 & 3.6\\
COSMOS-1   & 150.5067 &  2.7657  & 5.2 & 5.2 & 5.1\\
COSMOS-2   & 150.2063 &  1.8283  & 4.3 & 4.4 & 4.4\\
COSMOS-3   & 150.1838 &  1.6583  & 4.0 & 3.9 & 3.7\\
COSMOS-4   & 150.0987 &  2.7158  & 4.0 & 3.7 & 2.9\\
COSMOS-5   & 149.8486 &  2.2883  & 5.1 & 4.9 & 4.3\\
ELAIS-1    & 243.3380 & 55.4445  & 4.0 & 4.0 & 3.8\\
ELAIS-2    & 242.8025 & 55.5614  & 6.5 & 6.5 & 6.6\\
ELAIS-3    & 242.6855 & 55.4296  & 4.2 & 4.2 & 3.6\\
ELAIS-4    & 241.2850 & 55.6263  & 4.2 & 4.2 & 4.2\\
ELAIS-5    & 241.2278 & 54.5485  & 4.2 & 3.9 & 3.4\\
ELAIS-6    & 241.2034 & 54.4808  & 4.1 & 3.9 & 3.4\\
ELAIS-7    & 241.1764 & 54.5356  & 5.3 & 5.2 & 5.2\\
ELAIS-8    & 240.8272 & 55.1126  & 4.3 & 4.3 & 4.3\\
\hline
\end{tabular}
\end{table}

%
%%%%% Fig-A1
%
\begin{figure}
\begin{center}
 \includegraphics[height=88mm,angle=-90]{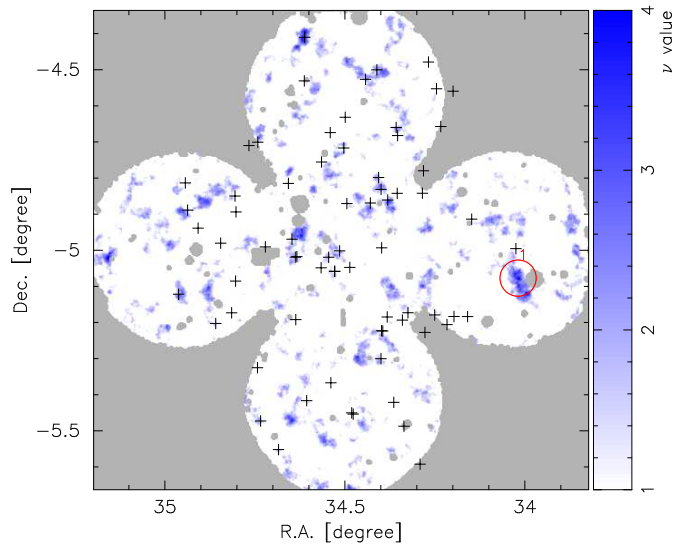}
\end{center}
\caption{$\cal K$ maps for the SXDS field, generated with the PEX0
  filter are plotted by 
  blue scale. Gray regions are either the region outside the analyzed
  area (within 15~arcmin from the field centre of Suprime-Cam) or
  masked regions where the data are affected by bright stars. 
  High peaks with $\nu>4$ are marked with red circle along with ID
  given in Table~ \ref{table:peaks}. The Plus symbols show positions of known
  clusters taken from a compilation by NASA/IPAC Extragalactic
  Database (NED). 
\label{fig:A1}}
\end{figure}

%
%%%%% Fig-A2
%
\begin{figure}
\begin{center}
 \includegraphics[height=88mm,angle=-90]{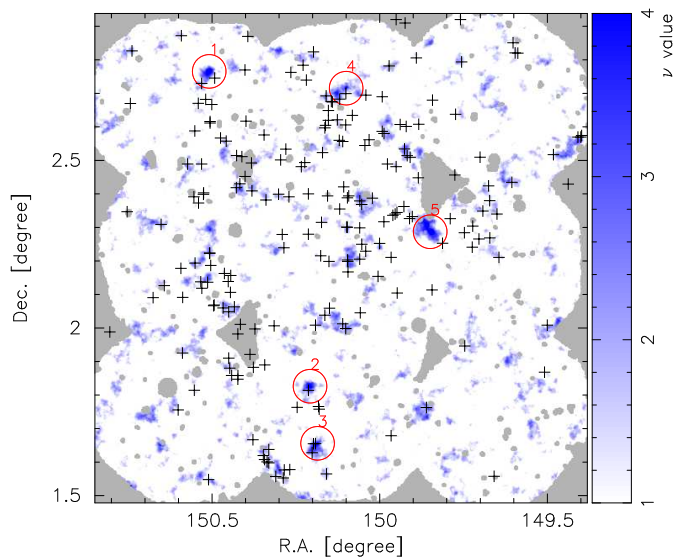}
\end{center}
\caption{Same as Fig.~\ref{fig:A1} but for the COSMOS field.
\label{fig:A2}}
\end{figure}

%
%%%%% Fig-A3
%
\begin{figure}
\begin{center}
 \includegraphics[height=88mm,angle=-90]{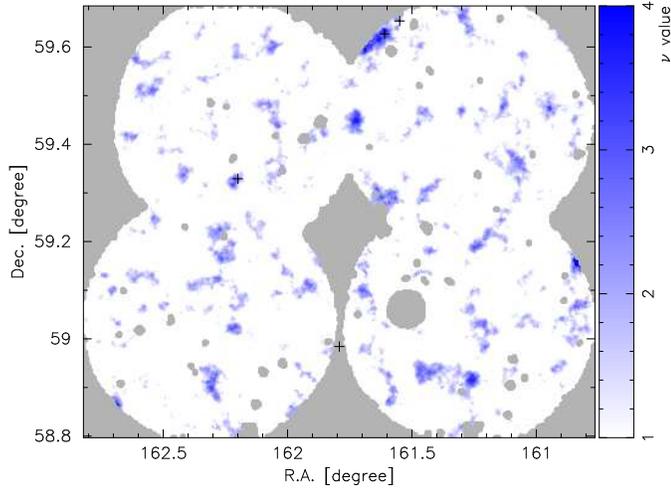}
\end{center}
\caption{Same as Fig.~\ref{fig:A1} but for the Lockman-hole field.
\label{fig:A3}}
\end{figure}

%
%%%%% Fig-A4
%
\begin{figure}
\begin{center}
 \includegraphics[height=88mm,angle=-90]{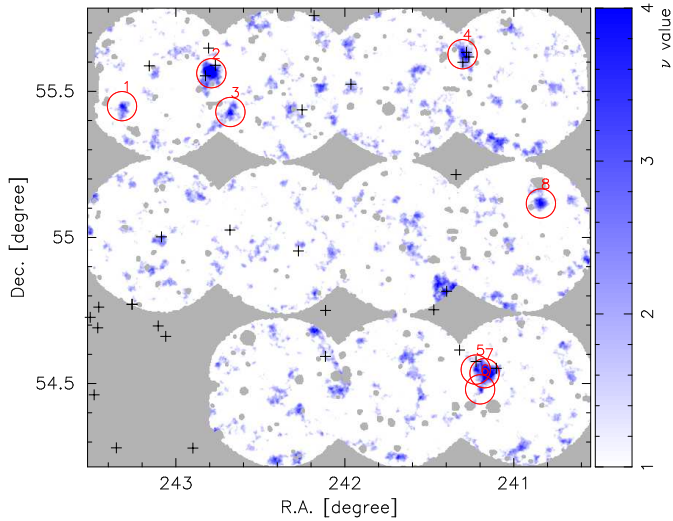}
\end{center}
\caption{Same as Fig.~\ref{fig:A1} but for the ELAIS N1 field.
\label{fig:A4}}
\end{figure}

Weak lensing $\cal K$ are computed from galaxy ellipticity (after the
PSF correction) data with three filters under consideration on regular
grid points with a grid spacing of 0.15~arcmin. We then identify high
peaks from the $\cal K$ maps. Peaks located within 1~arcmin from the
field boundary were discarded because such regions are likely affected
by the partial lack of data. The total area used for the peak finding
is 4~deg$^2$. We detect 14, 9 and 7 peaks with $\nu>4$ for the
PEX0, PEX1 and Gaussian filter, respectively.
High peaks with $\nu>4$ are summarized in Table \ref{table:peaks}, and
peak counts are plotted in Fig.~\ref{fig11.ps}, in which
a good agreement with mock simulation data is found.
In Figs.~\ref{fig:A1}-\ref{fig:A4}, $\cal K$ maps generated with the
PEX0 filter are shown in which high peaks are marked with circles. 
We also mark known clusters taken from a compilation by NASA/IPAC
Extragalactic Database (NED)\footnote{\tt
  http://ned.ipac.caltech.edu/} by the plus symbols. It is seen that many but
not all high peaks are associated with known clusters. It should be
noted that the known cluster sample plotted in the figures is just a
compilation of many individual cluster catalogues identified by many
different techniques, and thus is not homogeneous even over a single
field.  

\label{lastpage}
\end{document}